\documentclass[12]{article}
\usepackage{graphicx}
\usepackage{amssymb}

\begin{document}

\begin{center}
\Large{ Behaviour of a muonic atom as an acceptor centre in  diamond}

\vskip 5 mm

\large{T.N.~Mamedov$^{1}$, A.S.~Baturin$^{2}$,
K.I.~Gritsaj$^{1}$, A.~Maisuradze$^{3}$, V.G.~Ralchenko$^{4}$, 
R.~Scheuermann$^{5}$, K.~Sedlak$^{5}$, A.V.~Stoykov$^{5}$ }

\end{center}

\vskip 4 mm
\noindent
$^{1}$Joint Institute for Nuclear Research, 141980 Dubna,
                        Moscow reg., Russia\\
$^{2}$Moscow Institute of Physics and Technology, 141700 Dolgoprudny, 
           Moscow reg., Russia\\
$^{3}$Physik-Institut der Universit\"at Z\"urich, Winterthurerstrasse 190,
                 CH-8057 Z\"urich, Switzerland\\
$^{4}$Natural Sciences Center of Institute of General Physics, RAS,
           119991  Moscow, Russia\\
$^{5}$Paul Scherrer Institut, CH-5232 Villigen PSI,  Switzerland\\

\begin{abstract}
Polarized negative muons were used to study the behaviour of the boron
acceptor centre in synthetic diamond produced by the chemical vapour
deposition (CVD) method.  The negative muon substitutes one of the electrons
in a carbon atom, and this muonic atom imitates the boron acceptor impurity
in diamond.  The temperature dependence of the muon spin relaxation rate and
spin precession frequency were measured in the range of $20 - 330$~K in a
transverse magnetic field of 14~kOe.  For the first time a negative shift of
the muon spin precession was observed in diamond.  It is tentatively
attributed to an anisotropic hyperfine interaction in the boron acceptor.

The magnetic measurements  showed  that the magnetic susceptibility of the
CVD sample was close to that  of  the purest   natural diamond.

\end{abstract}


Diamond with its unsurpassed mechanical strength, thermal conductivity, and
radiation hardness is a promising semiconductor for particle detectors and
electronic components capable of withstanding high heat and radiation loads.
Great advances have been made over the last years in the technology of
manufacturing synthetic single crystal diamond and diamond
films~\cite{Isberg2002,Tuve2007,element6}.

Boron is the only dopant which forms an acceptor centre (AC) in diamond with
an ionization energy of $\approx 370$~meV \cite{Bezrukov1970}.  The
metal-insulator transition occurs at a concentration of $\approx 2\cdot
10^{20}$~cm$^{-3}$ of boron atoms \cite{Willams1970}.  The EPR signal of
boron impurities in diamond was observed only for uniaxially stressed
samples~\cite{Ammerlaan1980}, and the electronic state of this acceptor is
investigated insufficiently.

The possibility of using negative muons to study the behaviour of acceptor
impurities in diamond arises from the fact that capture of a negative muon
by a carbon atom results in the formation of a muonic atom $_\mu$B with an
electron shell that is analogous to that of the boron atom.  The evolution
of the polarization of $\mu^-$ in the 1s atomic state depends on the
interaction of the muon spin with the electron shell of the muonic atom and
on the interactions of this muonic atom (acceptor centre) with the crystal
lattice.  Providing that the electron shell of $_\mu$B has a nonzero
magnetic moment (paramagnetic $_\mu$B), there is a hyperfine interaction
between the muon and the electron shell.  The efficiency of the hyperfine
interaction depends on the relaxation rate $\nu$ of the magnetic moment of
 $_\mu$B and on the hyperfine interaction constant $A_{hf}$. According to
theoretical calculations~\cite{Gorelkin2000}, relaxation of the muon spin
and a paramagnetic shift of its precession frequency are expected for $\nu
\gg |A_{hf}|$.  Under the assumption of an isotropic hyperfine interaction,
the paramagnetic shift should be of positive sign and inversely proportional
to temperature.

The boron atom as AC in diamond may be in the  diamagnetic (B$^-$) or
paramagnetic (B$^-$ + $h$) state.  In the latter case a hole is localized in
vicinity of the AC.  At equilibrium the boron acceptor in diamond is
expected to be paramagnetic below $\approx 300$~K.  In earlier experiments
\cite{Mamedov2008} we observed no frequency shift of the muon spin
precession in diamond at the level of $5\cdot 10^{-3}$.  The goal of the
present experiment was to improve the accuracy of the muon spin precession
frequency measurement.

\section{Measurements}

The time-differential $\mu$SR experiment was performed with the upgraded ALC
instrument \cite{ALC} located at the $\pi$E3 muon beam line of the proton
accelerator of the Paul Scherrer Institute in Switzerland.  In the $\pi$E3
beam line the muons were "transversely" polarized -- the angle between the
muon spin and the muon momentum was $\sim 45^\circ$.  The polarization of
negative muons in synthetic diamond was studied in the magnetic field of
14~kOe in the temperature range of $20-330$~K.  One additional measurement
was performed in 3.0~kOe at 140~K.  The magnetic field at the sample was
parallel to the momentum of the incoming muon.

The polycrystalline diamond was produced by microwave plasma assisted
chemical vapour deposition (CVD) in $\rm CH_4$/$\rm H_2$ mixtures.  The
sample with the diameter $\approx$ 60~mm was grown on a silicon substrate
that was chemically etched after deposition~\cite{Ralchenko97}.  The main
impurities in the sample were hydrogen ($\approx$70~ppm) and nitrogen
($\approx$1.5~ppm).  Other impurities were less than 0.1~ppm.  The
cylinder-shaped sample D6 with the diameter 25~mm and height 1.3~mm
(thickness 0.427~g/cm$^2$) for the $\mu$SR study, and two bricks (samples A
and B) with the dimensions 3x1.5x1.5 mm$^3$ (weight 34~mg) for magnetic
measurements were cut using a laser.  In the $\mu$SR experiment 85\% of the
incoming muons stopped in the sample and 15\% in the copper holder.  As a
reference, a graphite sample (diameter 26 mm and thickness 0.66~g/cm$^2$)
was measured by $\mu$SR under the same conditions as the diamond sample.

The time distribution of electrons from the $\mu^- \rightarrow e^-$ decay
was fitted by the function
\begin{equation}
\begin{array}{r}
N(t)= N_0({\rm C})[1 + a \cdot P(t) ]
e^{-t/\tau(\rm C)}\\
     \hspace{6ex} +N_0({\rm Cu})\cdot e^{-t/\tau({\rm Cu})} +bg\\
    P(t)=P_0\cdot e^{-R\cdot t}\cdot {\rm cos}(\omega t +\varphi)
\end{array}
\label{polariz}
\end{equation}

\noindent
where $N_0$(C) ($N_0$(Cu)) is proportional to the number of muons stopped in
the sample (in the copper holder); $\tau$(C) ( $\tau$(Cu)) is the mean
lifetime of the muon in the 1s state of carbon (copper); $a$ is the
coefficient of asymmetry of the space distribution of electrons taking into
account the actual parameters of the spectrometer; $P_0$ is the muon
polarization in the 1s state at $t=0$; $R$ is the muon spin polarization
damping rate; $\omega$ and $\varphi$ are the frequency and the initial phase
of the muon spin precession in the magnetic field, and $bg$ is the
time-independent background.

The muon polarization $P_0$ in the 1s state of copper is close to zero due
to the hyperfine interaction of the muon spin with the spin of the nucleus
and due to very fast transitions between the levels $F_+=I+S_{\mu}$ and
$F_-=I-S_{\mu}$,
where $I$ is the spin of the copper nucleus and $S_{\mu}$ is the muon spin
(see \cite{Winston1963,Brewer1984,Mamedov1999}).

\section{Results and discussion}

The behaviour of the muon spin polarization in the reference graphite sample
was studied at temperatures 9.6, 50, 100, 200, and 300 K in a magnetic field
of 14 kOe.  It was observed that in graphite the muon spin polarization and
the frequency of the muon spin precession do not depend on temperature:
$[P_0(T)-P_0(300~{\rm K} )]/P_0(300~{\rm K})\lesssim 0.06\pm 0.08$ and
$[\omega(T)-\omega(300~{\rm K} )]/\omega(300~{\rm K}) \lesssim (6\pm4)\cdot
10^{-5}$ (see also~\cite{Mamedov1999}).  The relaxation of the muon spin in
graphite was not observed within the accuracy of the measurements ($R <
0.05$~MHz).

\begin{figure}
\begin{center}
\includegraphics{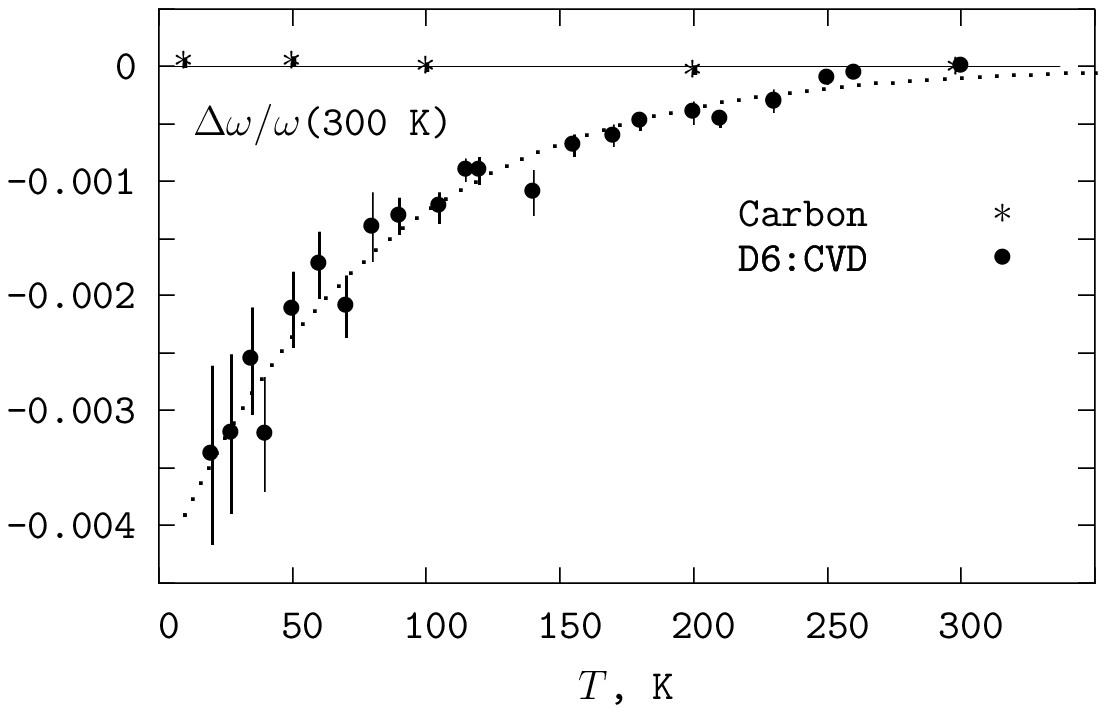}
\hspace{5ex}
\includegraphics{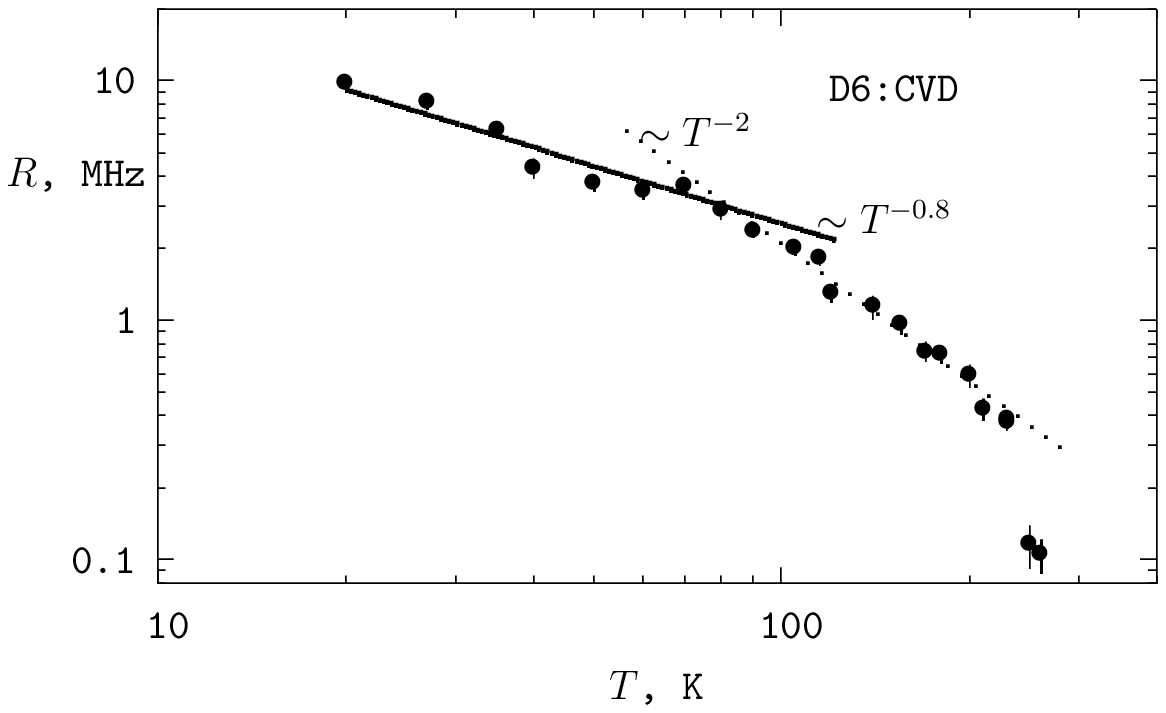}
\mbox{}\hspace{2em}\includegraphics{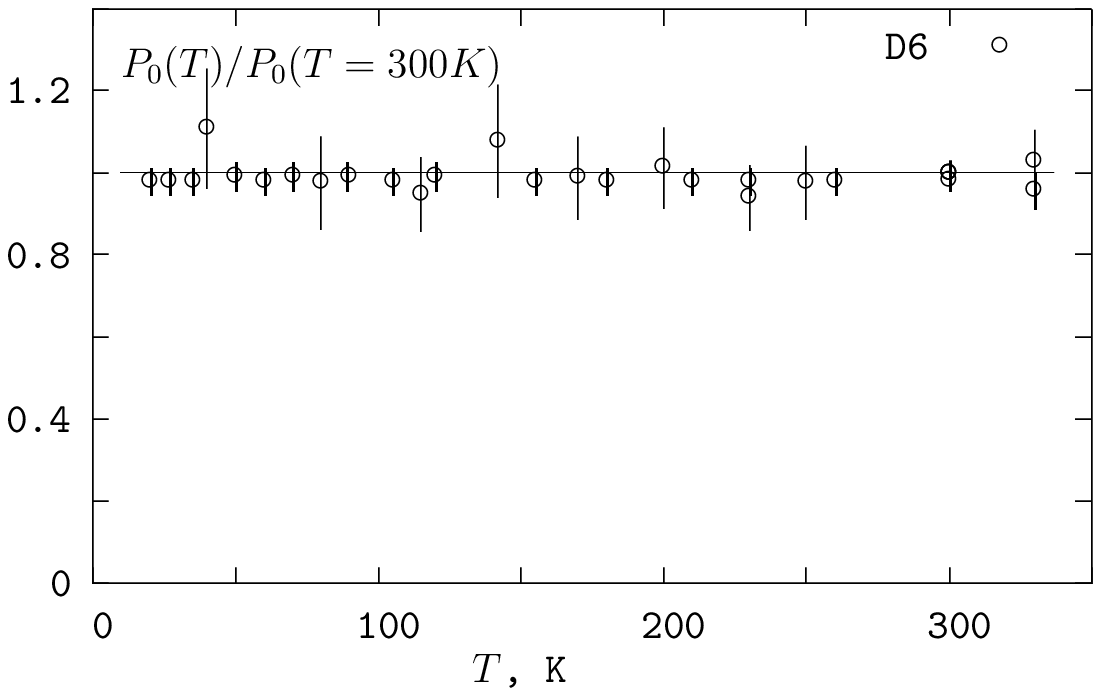}\\
\caption{Polarization amplitude $P_0(T)/P_0(T=300~{\rm K})$, damping rate
$R$ and precession frequency shift $\Delta \omega/\omega$ of the negative
muon spin measured in the synthetic CVD diamond sample D6 in the magnetic
field of 14 kOe.}
\label{D6-ALC-data}
\end{center}
\end{figure}

The results of the measurements on the CVD diamond  are shown in
Fig.~\ref{D6-ALC-data}.  In the sample D6 the muon polarization amplitude
$P_0$ does not depend on temperature and it is approximately 16~\% larger
than in the reference carbon sample (data with larger error bars are results
of preliminary measurements, the line is a guide to the eye).

The muon polarization damping rate depends on temperature as $R \sim
1/T^{0.8}$ in the range of $20--80$~K (solid line in Fig.~\ref{D6-ALC-data})
and as $R\sim 1/T^{2.0}$ in the range of $80-230$~K (dotted line).  A strong
drop of the damping rate at $\sim 250$~K and no damping at 300 and 330~K
were observed.  There is no difference in the damping rate of the muon spin
at 140~K in the magnetic field of 3.0~kOe compared to 14~kOe.  Generally,
the temperature dependence of $R$ in sample D6 is similar to that observed
earlier for the CVD film samples in 1.5 and 2.5~kOe magnetic
fields~\cite{Mamedov2008}.

As seen in Fig.~\ref{D6-ALC-data}, the muon spin precession frequency in the
carbon reference sample does not depend on temperature.  At room temperature
($T=300$~K) the muon spin precession frequency in the D6 sample is close to
that in the reference sample: $[\omega(C)-\omega(D6 )]/\omega(C) =
(1.8\pm0.4)\cdot10^{-4}$.  Opposite to the theoretical
prediction~\cite{Gorelkin2000} and to the experimental results in
silicon~\cite{Si-review}, a negative shift of the muon spin precession
frequency was observed at temperatures below 250~K.  The temperature
dependence of the frequency shift can be approximated as
$[\omega(T)-\omega(300~{\rm K})]/\omega(300~{\rm K}) = \Delta\omega/\omega
\sim -1/T$ (dotted line in Fig.~\ref{D6-ALC-data}).

The negative shift of the muon spin precession frequency  may be due to
three different reasons:

\noindent
a)  it  may be due to  the presence of other paramagnetic defects
(excluding the acceptor centres) in diamond.  The magnetic field at the
muon could  differ from the external field because of polarization of these
paramagnetic defects.  For example, N$^0$ and H1(H2) paramagnetic centres
were observed in polycrystalline CVD diamond by EPR~\cite{Ralch2000};

\noindent
b) let us assume that the muonic atom is formed in the $_{\mu}$B$^0$ state
within a time less than 10$^{-9}$~s and that the transition rate from this
state to the equilibrium one ($_{\mu}$B$^{-}$+h) is smaller than the muon
decay rate.  In the state $_{\mu}$B$^0$ the muonic atom has covalent
chemical bonds with three nearest carbon atoms, and the fourth carbon atom
has an electron with an unpaired spin (one dangling bond).  The polarization
of this paramagnetic complex in an external magnetic field could result in
the negative shift of the muon spin precession frequency;

\noindent
c) negative frequency shift can also be due to polarization of  the
paramagnetic acceptor centre in the magnetic field if the muonic atom is
formed in the ($_{\mu}$B$^{-}$+h) state in $\lesssim 10^{-8}$~s and the
hyperfine interaction in the AC is anisotropic.  In the state
($_{\mu}$B$^{-}$+h) the muonic atom has covalent chemical bonds with all
four nearest carbon atoms and a hole localized in the vicinity of AC.

The frequency shift of  the muon spin precession should be proportional to
the susceptibility of the paramagnetic defects or to the susceptibility of
the paramagnetic acceptor centres if $\nu \gg | A_{hf}|$.  Therefore, a
simple $1/T$ (Curie type) temperature dependence of the muon spin precession
frequency shift is expected.

\begin{figure}
\begin{center}
\includegraphics[width=\columnwidth]{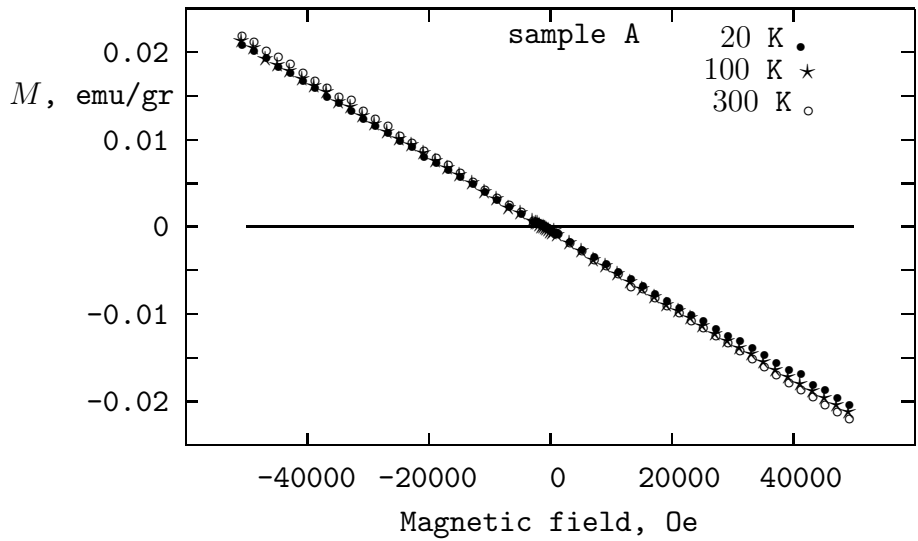}
\caption{Magnetic moment of  CVD diamond sample A vs the magnetic field
at the temperatures 20, 100, and 300~K.}
\label{squid1}
\end{center}
\end{figure}

\begin{figure}
\begin{center}
\includegraphics[width=\columnwidth]{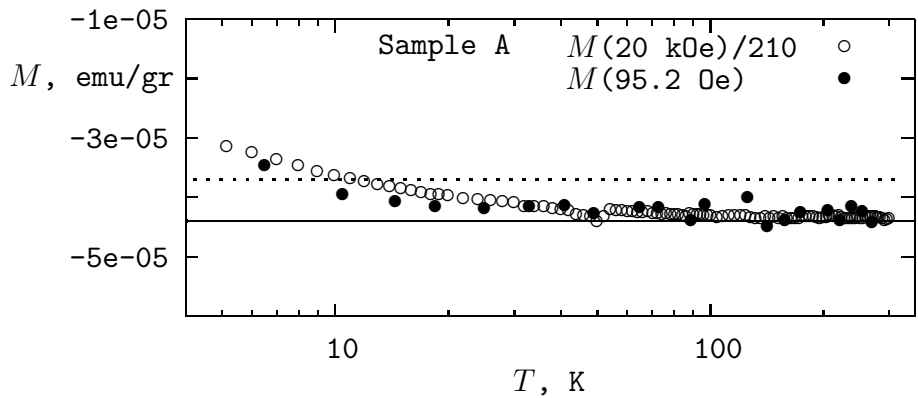}
\caption{Temperature dependence of the magnetic moment of the synthetic
CVD diamond sample A in the magnetic field of 95.2~Oe and 20.0~kOe.
For better visualization in one single graph
the value of $M$ at the magnetic field of 20.0~kOe was scaled by 1/210}
\label{squid2}
\end{center}
\end{figure}

To elucidate the effect of paramagnetic (magnetic) defects on the muon spin
precession frequency we measured the magnetic susceptibility of our sample.
The magnetic moments ($M$) of samples A and B were measured in magnetic
fields in the range of $0-50$~kOe at the temperatures 20, 100, and 300~K.
The temperature dependence $M(T)$ was studied in detail in the magnetic
fields of 95.2~Oe and 20.0~kOe for both samples.  The DC magnetization
measurements were performed with the Quantum Design SQUID magnetometer.  The
results for both samples A and B are similar within the accuracy of the
measurements.  The results of the measurements for sample A are presented in
Fig.~\ref{squid1} and in Fig.~\ref{squid2}.  There is evident linear
dependence of the magnetic moment on the magnetic field in the range from 0
to 50.0 kOe.  The susceptibility of the sample was $\chi=-4.13(2)\cdot
10^{-7}$~cm$^3$/g, $-4.29(2)\cdot 10^{-7}$~cm$^3$/g, and $-4.36(2)\cdot
10^{-7}$~cm$^3$/g, at the temperatures 20, 100, and 300 K, respectively.
The value of $\chi$ for the synthetic CVD sample is in agreement with that
for natural diamond~\cite{Heremans94,Yelis}.  As seen from
Fig.~\ref{squid2}, the magnetic moment of the sample in the 95.2 Oe and
20.0~kOe magnetic field practically does not depend on temperature in the
range of $20-300$~K.  Only at temperatures below 20~K the magnetic moment
slightly increases with decreasing temperature.  Therefore, the
concentration of magnetic impurities (defects) in our CVD sample is
negligible and cannot produce the observed frequency shift.  Moreover, the
muon spin precession frequency shift depends on temperature as $\sim -1/T$
in the range of $20-300$~K, where the magnetic moment of the sample is
practically constant.

The measured values of $P_0$ at different $T$ are close to the maximum
possible muon polarization in the 1s state of a carbon atom.  This means
that the muonic atom $_{\mu}$B is formed in one of the chemically bonded
$_{\mu}$B$^0$ or ($_{\mu}$B$^{1-}$+ h) states with a probability close to
unity in less than $10^{-9}$~s.  It is worth mentioning that the Auger
transitions of the muon destroy chemical bonds of the muonic atom with
neighbouring carbon atoms.
Theoretical calculations~\cite{Baturin2006,Baturin2007} suggest that the
muonic atom in diamond restores chemical bonds and is formed in the
$_{\mu}$B$^0$ state within $10^{-10} - 10^{-9}$~s.

At present we do not have an analytical expression for the relaxation of the
muon spin if there is an anisotropic hyperfine interaction in the muonic
atom.  Nevertheless, we may assume that $R$ is inversely proportional to the
relaxation rate of the magnetic moment of the acceptor centre in both
$_{\mu}$B$^0$ and ($_{\mu}$B$^{1-}$ +h) states.  In diamond relaxation of
the magnetic moments of $_{\mu}$B$^0$ and ($_{\mu}$B$^{-}$ +h) should be
attributed to scattering of phonons by these centres.  Direct, Raman, and
Orbach processes can contribute to phonon scattering, and the temperature
dependence of the probability of these processes is very
different~\cite{Yafet}.  In the present experiment two temperature domains
with different temperature dependences of the relaxation rate of the muon
spin were observed.  Thus, we conclude that there is a contribution of two
different processes to scattering of phonons at the acceptor in diamond.
The dominance of one process over the other reverses at about 80~K.
According to the calculations for the acceptor in silicon, the Raman process
overcomes the direct (one phonon) process at $T= 8$ K, and the Orbach
process is negligible below 100~K \cite{Yafet}.

\section{Conclusion}
 A large negative shift of the precession frequency of the
negative muon spin in diamond was found for the first time.  Simple Curie
type $\Delta\omega/\omega \sim -1/T$ temperature dependence of the frequency
shift was observed.  It is not excluded that the muonic atom as an AC forms
in the equilibrium ($_{\mu}$B$^{-}$ +h) state in $10^{-8}$~s and there is an
anisotropic hyperfine interaction in AC.  The temperature dependence of the
relaxation rate of the muon spin evidences that two different processes
contribute to phonon scattering by the acceptor centre in the temperature
range of $20-250$~K.

The magnetic measurements  showed  that the magnetic susceptibility of the
CVD sample was close to that of the purest natural diamond.

The authors are grateful to prof. Yu.M.Belousov for fruitful discussions. 
We thank the Directorate of the Paul Scherrer Institute for giving us the
possibility of carrying out the experiments at PSI.



\begin{thebibliography}{99}
\bibitem{Isberg2002} J.Isberg et al., Science {\bf 297} (2002) 1670.

\bibitem{Tuve2007} C.Tuve  et al., Nucl.Instr. and Meth. Phys.
                      Res.~A {\bf 570} (2007) 299.

\bibitem{element6} {\tt http://www.e6.com} (26.06.2012)

\bibitem{Bezrukov1970} G.N.Bezrukov  et al.,
                       Sov. Phys. Semicond. {\bf 4} (1970) 587.

\bibitem{Willams1970} A.W.S.Willams, E.C.Lightowlers, A.T.Collins,
                      J.Phys. C {\bf 3} (1970) 1727.

\bibitem{Ammerlaan1980} C.A.J.Ammerlaan,
                         Inst. Phys. Conf. Series~{\bf 59} (1980) 81.

\bibitem{Gorelkin2000} V.N.Gorelkin, T.N.Mamedov, A.S.Baturin,
                      Physica~B  {\bf 289-290} (2000) 585.

\bibitem{Mamedov2008} T.N.Mamedov, A.S.Baturin, V.D.Blank et al.,
                    Diamond Relat. Matter. {\bf 17} (2008) 1221-1224.

\bibitem{ALC} {\tt http://lmu.web.psi.ch/facilities/alc/alc.html} (26.06.2012)

\bibitem{Ralchenko97} V.G.Ralchenko et al.,
                        Diamond Relat. Mater. {\bf 6} (1997) 417.

\bibitem{Winston1963} R.Winston, Phys.Rev. {\bf 129} (1963) 2766.

\bibitem{Brewer1984} J.H.Brewer, Hyperfine Interaction {\bf 19} (1984) 873.

\bibitem{Mamedov1999} T.N.Mamedov et al.,
                       J.Phys:~Condens.Matter {\bf 11} (1999) 2849.

\bibitem{Ralch2000} S.N.Nistor, M.Stefan, V.Ralchenko et al.,
                  J.Appl. Phys. {\bf 87} (2000) 8741.

\bibitem{Si-review}T.N.{\,}Mamedov, V.N.{\,}Gorelkin,  A.V.{\,}Stoykov,
                    Physics Particle Nuclei {\bf 33}  (2002) 519.

\bibitem{Heremans94} J.Heremans, C.H.Olk, D.T.Morelli,
                     Phys.Rev. B {\bf 49} (1994) 15122.

\bibitem{Yelis} A.P.Yelisseyev,V.P.Afanasiev, V.N.Ikorsky,
                Doklady Earth Sciences {\bf 425(2)} (2009) 330-333.

\bibitem{Baturin2006} A.S.Baturin, V.N.Gorelkin, V.S.Rastunkov,
                     V.R.Soloviev, Physica B {\bf 374-375} (2006) 340.

\bibitem{Baturin2007} A.S.Baturin, V.N.Gorelkin, T.N.Mamedov et al.,
               JINR communucation, P14-2007-188, Dubna, 2007, Russia (in russian).

\bibitem{Yafet} Y.Yafet, J.Phys.Chem.Solids {\bf 26} (1965) 647.

\end{thebibliography}
\end{document}